\documentclass{article}
\usepackage{microtype}
\usepackage{graphicx}
\usepackage{booktabs}
\usepackage{array}
\usepackage{tabularx}
\newcolumntype{R}{>{\raggedleft\arraybackslash}X}
\newcolumntype{C}{>{\centering\arraybackslash}X}
\usepackage{amsmath,amssymb}
\usepackage{xurl}
\usepackage{placeins}
\usepackage{needspace}
\usepackage{flafter}
\usepackage{hyperref}
\usepackage{xcolor}
\usepackage{colortbl}
\definecolor{tableheadergray}{HTML}{E8E8E8}

\IfFileExists{paper2/arxiv/mlsys2025.sty}{\def\paperpath{paper2/arxiv/}}{%
  \IfFileExists{arxiv/mlsys2025.sty}{\def\paperpath{arxiv/}}{\def\paperpath{}}}
\usepackage[accepted]{\paperpath mlsys2025}
\makeatletter
\renewcommand{\Notice@String}{Preprint.}
\makeatother
\hypersetup{
  pdftitle={Understanding the Energy Cost of Quantized Agents on Shared GPUs},
  pdfauthor={Stephen Abkin, Alex Flesher, Aldric Yang, Nitheesh Kodarapu, Deepak Panigrahy, David Kebo Houngninou},
  pdfsubject={Energy measurement for quantized language-model agents on shared GPUs},
  pdfkeywords={language-model agents, quantization, inference serving, energy measurement}
}
\providecommand{\doi}[1]{doi: \href{https://doi.org/#1}{\begingroup\urlstyle{rm}\nolinkurl{#1}\endgroup}}
\mlsystitlerunning{Understanding the Energy Cost of Quantized Agents on Shared GPUs}

\usepackage{enumitem}
\usepackage{tcolorbox}
\definecolor{takeawaygreen}{HTML}{D2E6D2}
\definecolor{takeawaywash}{HTML}{F3F8F2}
\definecolor{takeawayedge}{HTML}{AECBAE}
\newtcolorbox{paperinsight}[1]{colback=takeawaywash,colframe=takeawayedge,
 colbacktitle=takeawaygreen,coltitle=black,fonttitle=\bfseries,
 title={#1},boxrule=0.45pt,arc=1.5pt,left=4pt,right=4pt,
 top=3pt,bottom=3pt,toptitle=2pt,bottomtitle=2pt,
 before skip=5pt,after skip=5pt}
\newcommand{\takeaway}[2]{\begin{paperinsight}{Takeaway #1}#2\end{paperinsight}}

\begin{document}
\raggedbottom%
\twocolumn[
\mlsystitle{Understanding the Energy Cost of\texorpdfstring{\\}{ }Quantized Agents on Shared GPUs}
\mlsyssetsymbol{equal}{*}
\begin{mlsysauthorlist}
\mlsysauthor{Stephen Abkin}{tamu}
\mlsysauthor{Alex Flesher}{equal,tamu}
\mlsysauthor{Aldric Yang}{equal,tamu}
\mlsysauthor{Nitheesh Kodarapu}{tamu}\par
\mlsysauthor{Deepak Panigrahy}{tamu}
\mlsysauthor{David Kebo Houngninou}{tamu}
\end{mlsysauthorlist}
\mlsysaffiliation{tamu}{Texas A\&M University, College Station, Texas, USA}
\mlsyscorrespondingauthor{Stephen Abkin}{\href{mailto:stephen122204@tamu.edu}{stephen122204@tamu.edu}}
\mlsyskeywords{language-model agents, quantization, inference serving, energy measurement}
\vskip 0.3in
\begin{abstract}
Quantization can lower the energy of language-model calls and free memory for serving more agents, but its benefit per completed task remains unclear when agents share a GPU. Measurements of fixed requests or isolated agents leave open how the agents' generated work interact with competition for compute and cache. We present a systematic study of precision and admission limits for two agent models on A100 deployments with 40 or 80 GB of memory, using coding and data-analysis tasks with independently verified outcomes. We replay recorded calls across precisions to isolate execution energy and cap the cache to test the value of freed memory. For both models with four admitted agents where requests fit in the cache, 4-bit weights with 16-bit activations execute the same calls with 10–25\% less energy than 8-bit weights and activations. Despite this saving, extra work and fewer completions in live runs raise their energy per verified completion by 49\%. Where requests compete for cache on the 40 GB deployment, 4-bit weights cost less, and capping their cache raises energy per verified completion by 34\% with 16 agents. These results show how generated work and cache demand change the energy benefit of quantization, providing a practical basis for choosing precision and admission limits together.
\end{abstract}
]
\printAffiliationsAndNotice{\textsuperscript{*}Alex Flesher and Aldric Yang contributed equally. }
\setcounter{footnote}{1}
\interfootnotelinepenalty=10000

\section{Introduction}
Independent language-model agents work through software repairs and data-analysis tasks, each alternating model responses with tools that query data, edit programs, or run tests. Open-weight coding models let an operator host these agents on infrastructure it controls~\cite{mistral2025devstral11}. Serving the agents on a shared GPU requires choosing both the model's precision and an \emph{admission limit} $N$, the number of sessions allowed to run concurrently. We compare each combination, or \emph{setting}, by the energy needed per verified completion in live runs (i.e., the actual task attempts). Energy spent on unsuccessful attempts contributes to the cost of the tasks the service does complete, so cheaper execution is useful only if enough tasks finish correctly and within the service's latency requirement.

Quantization offers two potential benefits for this service. Lower-precision weights can reduce the energy of each model call and leave more GPU memory for reusable context stored as attention keys and values in the key--value (\emph{KV}) cache. Lower precision can also change an agent's responses, which shape the work that follows. A response may need correction, a rejected tool request may be repeated, or a long attempt may end without a correct solution. Each extra step consumes energy, so a saving on individual calls can be offset by the calls needed to finish a task.

Energy studies measure inference energy per output token~\cite{chung2026joules} or per response~\cite{chung2025mlenergy}, with savings that depend on the kernels and workload~\cite{MLSYS2025_fbe2b2f7,chung2026joules}. Throughput and energy can also favor different settings~\cite{chung2025mlenergy,chung2026joules}. These measurements can suggest a setting to deploy, but they do not directly count how many agent tasks the service completes. Related studies of cache compression show that longer responses can offset a throughput gain~\cite{gao2025rethinkingkv}. For one agent on a dedicated GPU, measurements of quantized agents connect energy per completion to both energy per token and tokens generated per completion~\cite{abkin2026taskenergy}.

Sharing a GPU adds competition for compute and cache to this relationship between execution cost and generated work. Admitting more agents creates opportunities to batch their calls, but their contexts also compete for memory. Serving studies connect performance to memory management, quantized kernels, and workload shape~\cite{kwon2023pagedattention,MLSYS2025_fbe2b2f7,MLSYS2026_8144a9d6}, while multi-request measurements expose the energy cost of dependencies between calls~\cite{ifath2026multirequest}. Memory freed by quantization may help admitted requests fit in the cache, even when a change in precision increases the work agents generate. Single-response and single-agent measurements do not show how these effects combine in a shared service. An operator choosing by energy per token or throughput may select a setting that is cheaper per call but costlier per verified completion.

\begin{figure}[t]
\centering
\includegraphics[width=\columnwidth,trim=0 5.3bp 0 0.5bp,clip]{\paperpath 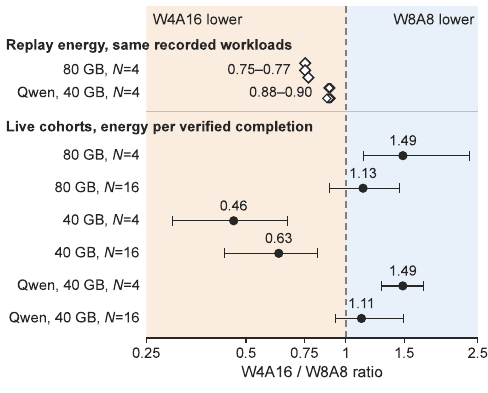}
\caption{W4A16 saves replay energy but costs more per verified completion with four agents where requests fit in the cache. Diamonds show replay-energy ratios; circles show live ratios of energy per verified completion. Bars show 95\% intervals from paired repetitions (Devstral: six on 80 GB, seven on 40 GB; Qwen: 11 at $N=4$, four at $N=16$). Values below one favor W4A16. $N$ is the admission limit; rows not labeled Qwen use Devstral.}
\label{fig:preview}
\end{figure}

\begin{samepage}
\noindent\textbf{We study how precision and admission limits jointly affect energy per verified completion when agents share a GPU.} We investigate this relationship through four research questions: \par
\end{samepage}

\begin{enumerate}[leftmargin=2.6em]
\Needspace{4\baselineskip}
\item[\textbf{RQ1:}] Does the precision with the lowest energy per verified completion depend on the admission limit and the GPU and serving configuration?
\Needspace{4\baselineskip}
\item[\textbf{RQ2:}] Does a saving on fixed recorded model calls carry through to the energy per verified completion of live agents?
\Needspace{4\baselineskip}
\item[\textbf{RQ3:}] Does memory freed by quantization lower energy per verified completion when requests compete for cache?
\Needspace{4\baselineskip}
\item[\textbf{RQ4:}] Do settings selected in preliminary (pilot) runs for low energy per token or high token throughput meet the service's completion and latency requirements?
\end{enumerate}

We compare three precisions, BF16, W8A8, and W4A16, for two agent models, Devstral Small 1.1 and Qwen3.5-9B.\footnote{BF16 is the 16-bit bfloat16 format. In W8A8 and W4A16, W gives the bit width of stored weights and A gives the bit width of activations, the intermediate values computed during inference. W8A8 uses 8-bit integer weights and activations, and W4A16 uses 4-bit integer weights and BF16 activations.} Their published baseline and quantized checkpoints let us compare deployable precisions of the same trained model. Devstral runs on 80 GB and 40 GB A100 GPUs, and Qwen runs on the 40 GB GPU, where all three precisions fit. For \textbf{RQ1}, we compare energy per verified completion across precisions and admission limits on a shared vLLM server~\cite{kwon2023pagedattention}. %
Each live run offers 16 data-analysis and repository-repair tasks with independently checked outcomes.

To explain the live comparisons, we replay recorded calls across precisions to isolate the energy of executing the same work (\textbf{RQ2}). We cap the KV cache to test whether freed memory lowers energy per verified completion (\textbf{RQ3}). For \textbf{RQ4}, we check whether two token-based rules fixed before the main 80 GB Devstral runs choose settings that meet completion and latency requirements.

\paragraph{\upshape\bfseries Summary of findings.} On 80 GB, the precision with the lowest observed cost changes with admission. On 40 GB, Devstral's W4A16 costs less than W8A8 at both tested limits (Section~\ref{sec:outcomes}). In our comparisons with four admitted agents where requests fit in the cache, extra work and fewer completions raise W4A16's energy per verified completion by 49\% over W8A8, despite its saving on fixed recorded calls (Figure~\ref{fig:preview} and Section~\ref{sec:work}). Where requests compete for cache, freed memory contributes to a lower cost. Capping Devstral's W4A16 cache on 40 GB raises its energy per verified completion by 34\% with 16 agents (Section~\ref{sec:capacity}). On the 80 GB Devstral deployment, settings selected by token measurements from the preliminary (pilot) runs fail both the completion and latency requirements (Section~\ref{sec:selection}). These results support evaluating precision and admission together on the intended tasks, checking service requirements before comparing energy per verified completion.

We make three concrete contributions:
\begin{itemize}
\item \textbf{Task-level measurement of precision and admission on a shared GPU.} We measure energy per verified completion, including failed attempts, for three precisions at up to four admission limits across two models and two A100 deployments (Sections~\ref{sec:metric} and~\ref{sec:design}).
\item \textbf{Replay of recorded agent work across precisions.} We separate execution savings from differences in generated work by replaying each recorded workload across precisions (Section~\ref{sec:work}).
\item \textbf{Testing the energy benefit of memory freed by quantization.} We test the value of freed memory by capping W4A16's KV cache at W8A8's capacity within paired repetitions (Section~\ref{sec:capacity}).
\end{itemize}

\section{Background and Related Work}
\label{sec:background}
\paragraph{Execution and cache capacity.}
QServe combines quantization algorithms with execution kernels to make low-precision serving efficient~\cite{MLSYS2025_fbe2b2f7}. The server must also manage requests as their contexts occupy memory. Orca allows requests to join or leave a batch between model iterations~\cite{yu2022orca}, and PagedAttention manages cache allocation and sharing~\cite{kwon2023pagedattention}. When active requests need more cache than is available, the server preempts a request and later recovers its state by recomputation or swapping~\cite{kwon2023pagedattention}. This recovery adds work or data movement to serving. MorphServe coordinates precision and cache capacity at runtime by swapping in quantized layers and resizing the KV cache under load~\cite{MLSYS2026_8144a9d6}. Freed memory can reduce recovery work when admitted requests exceed the cache.

\paragraph{Requests within an agent task.}
A returning agent sends its conversation and tool output to the model. Prefix caching can reuse the unchanged beginning of this input, while tool activity determines when the next request arrives. Parrot exposes dependencies between requests and shared prompt text so that a server can optimize an application across calls~\cite{lin2024parrot}. Once requests arrive, scheduling determines how their prompt processing and token generation share the GPU. Sarathi-Serve processes prompts in chunks alongside token generation to change the tradeoff between throughput and latency~\cite{agrawal2024sarathi}.\footnote{Generating tokens after the prompt is called decoding. Throughput counts tokens or requests processed per unit time, while latency measures their elapsed time.} For speculative decoding, \citet{MLSYS2026_554e056f} relate speedup to execution costs and token acceptance. For agents, we need to account for the energy of successive calls and whether the task is completed successfully.

\paragraph{Energy and verified outcomes.}
ML.ENERGY characterizes inference energy across serving configurations and workloads~\cite{chung2025mlenergy}, and \citet{chung2026joules} show that lower precision can increase energy at small batch sizes. Measurements of multi-request workflows count energy across sequences of calls~\cite{ifath2026multirequest}. Evaluating the outcome of those calls also requires attention to the agent's tools. SWE-agent shows that the tool interface can change behavior and task success~\cite{yang2024sweagent}, and agent-evaluation work argues for considering accuracy and cost together~\cite{kapoor2025agents}. Energy per successful task is used to evaluate quantized agents~\cite{abkin2026taskenergy}. \citet{siraj2026smaller} model energy, monetary cost, and time per successful task to study how model selection changes with task length. For concurrent agents, the accounting problem is to relate verified task outcomes to energy consumed jointly by overlapping sessions.

\section{Energy per Verified Completion}
\label{sec:metric}
\begin{figure*}[t]
\centering
\includegraphics[width=\textwidth]{\paperpath 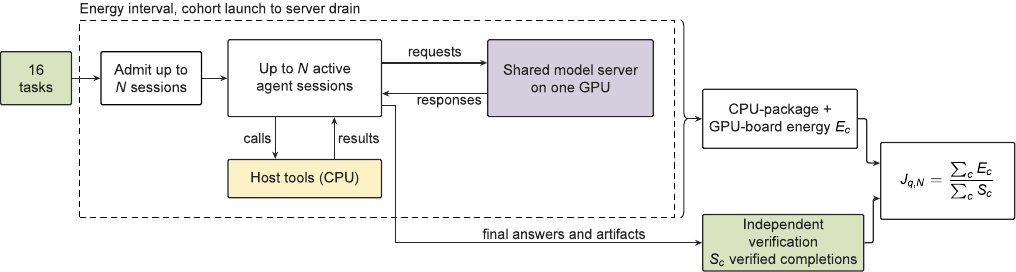}
\caption{Energy per verified completion pools each cohort's energy and independently checked outcomes. CPU-package and GPU-board energy over the dashed interval, from cohort launch through admission to server drain, give $E_c$. Verification afterward gives $S_c$, and Equation~\ref{eq:cost} pools these quantities across cohorts.}
\label{fig:design}
\end{figure*}

Failed attempts consume energy even when they produce no completed task, so we measure the energy of a whole cohort of tasks offered together and count its independently verified completions. Each task receives an independent agent session, and the admission limit $N$ sets how many sessions may run at once.
A waiting task is admitted when a running session ends. An admitted session may be running a tool rather than waiting on the model, so the number of active model requests is often below $N$. Figure~\ref{fig:design} shows how the cohort's energy and independently verified outcomes enter the measurement.

Let $E_c$ be the CPU-package plus GPU-board energy of cohort $c$, measured from cohort launch until all pending server requests finish (server drain). Let $S_c$ count the tasks that pass independent verification. For precision $q$ and admission limit $N$, we pool energy and successes over the evaluated cohorts,
\begin{equation}
 J_{q,N}=\frac{\sum_c E_c}{\sum_c S_c}.
 \label{eq:cost}
\end{equation}
This energy includes failed attempts, watchdog stops (i.e., sessions ended at the 1,800-second time limit), tasks that end normally but fail verification, and the final period in which one session runs alone. Measuring the whole cohort counts the shared server once, whereas summing energy over agents' overlapping time windows would double-count some energy. We call $J_{q,N}$ the \emph{cost of a setting}, defined only when at least one task is verified. By contrast, energy per output token divides the same energy by generated tokens, so a precision whose agents generate more tokens to reach the same outcomes can look cheaper per token while costing more per verified completion. 

\Needspace{8\baselineskip}
At a fixed admission limit within one deployment, switching precision can change both total energy and the number of successes. Let $r_E=E_{\mathrm{W4}}/E_{\mathrm{W8}}$ compare the total energy of W4A16 with that of W8A8, and let $r_S=S_{\mathrm{W4}}/S_{\mathrm{W8}}$ compare their verified completions. Both totals sum over the same paired repetitions. Applying Equation~\ref{eq:cost} gives
\begin{equation}
 \frac{J_{\mathrm{W4}}}{J_{\mathrm{W8}}}=\frac{r_E}{r_S},
 \qquad J_{\mathrm{W4}}<J_{\mathrm{W8}}\quad\Longleftrightarrow\quad r_E<r_S.
 \label{eq:boundary}
\end{equation}
This accounting identity expresses the tradeoff between energy and completed work~\cite{abkin2026taskenergy,siraj2026smaller} for agents sharing one measurement interval. For example, if W4A16 uses $0.8$ times the energy of W8A8, it costs less per verified completion only if it completes more than $0.8$ times as many tasks.

\section{Experimental Design}
\label{sec:design}
Live runs compare precision and admission (\textbf{RQ1}). Replay and cache caps test execution savings and freed memory (\textbf{RQ2--RQ3}), while token-based rules check whether selected settings meet service requirements (\textbf{RQ4}).

\paragraph{Model and precision choices.}
We use Devstral Small 1.1, a 24-billion-parameter model designed for software-engineering agents, to compare deployable precisions of the same trained model. Its original BF16 weights and matched Red Hat W8A8 and W4A16 checkpoints provide that comparison~\cite{mistral2025devstral11,redhat2025devstralw4a16,redhat2025devstralw8a8}.\footnote{Mistral AI and All Hands AI developed Devstral Small 1.1, published as \texttt{Devstral-Small-2507}. A checkpoint contains saved model weights and configuration files.} The later Devstral Small 2 Instruct checkpoint is published in FP8, so Small 1.1 supplies the BF16 reference needed here~\cite{mistral2025devstralsmall2}. The quantized checkpoints use GPTQ~\cite{frantar2023gptq}, with SmoothQuant for Devstral's W8A8~\cite{xiao2023smoothquant}. A difference between two precisions measures the combined effect of the formats, quantization recipes, and kernels that an operator would deploy together.

Within each model, precision comparisons share the source model, tokenizer, prompts, tools, and sampling settings.\footnote{A tokenizer converts text into model input units called tokens. Sharing it keeps token counts comparable across precisions.} All three Devstral precisions retain a BF16 KV cache. Sampling follows the publisher's recommendation, temperature 0.15 and top-$p=1$. Supplementary Section A gives full configurations and paired tests.

\paragraph{Shared serving and memory capacity (RQ1).}
The main experiment varies admission on an NVIDIA A100 PCIe 80 GB GPU, where all three Devstral precisions fit. To compare the quantized versions under tighter memory, we add a second deployment on an A100 PCIe 40 GB GPU, where their weights fit and those of BF16 do not. Both Devstral deployments use vLLM 0.10.2~\cite{kwon2023pagedattention}, reserve 90\% of GPU memory for serving, and enable prefix caching to reuse the unchanged beginning of a returning request. The context limits are 131,072 tokens on 80 GB and 65,536 tokens on 40 GB. On 80 GB, BF16, W8A8, and W4A16 provide measured cache capacities of 176,368, 311,552, and 377,120 tokens, respectively. Because the machines and context limits also differ, we analyze the deployments separately and draw causal conclusions only from interventions within each deployment.

\paragraph{Workloads and independent verification.}
The 16-task catalogue combines structured data operations with repository repairs that require inspecting, editing, and testing code. Eight data-analysis tasks cover SQL analytics, grouped regression, extract--transform--load (ETL) pipeline repair, and Python data processing, with two tasks per family at different input sizes. Eight issues from SWE-bench Verified~\cite{jimenez2024swebench,openai2024swebenchverified} span five repositories and three human-annotated difficulty levels. The data tasks use templates tested in the pilot runs (i.e., agent runs made before the experiment to select templates and admission limits), with fresh inputs and grading challenges, and the repository issues are disjoint from the pilot instances. The catalogue includes both short tasks and longer attempts with repeated tool use. With BF16 at four admitted agents on 80 GB, data-analysis attempts take a median of three to five turns per family, while repository attempts take a median of 31.5 turns, and 10 of 48 repository attempts reach 250 turns (Supplementary Section A, Table S2).

A task passes only when an independent verifier accepts its final answer or artifact. Before the experiment, we checked each verifier against a reference solution and an unchanged submission. Each agent has its own conversation and workspace. Repository agents run in sanitized containers with networking disabled, and the task descriptions do not reveal the verifier's checks. Tests requested by an agent remain part of its measured work, while independent verification runs after the energy interval.

\paragraph{Agent execution and stopping rules.}
Each model response counts as one turn. A requested tool call executes if it meets the declared schema, which specifies the function name, arguments, and types. Otherwise, the agent receives an error and can revise its request. A session may use up to 250 turns (its turn budget) and 500 tool calls. A watchdog stops it after 1,800 seconds, %
almost three times the longest observed verified completion. These limits follow the expanded budget of \citet{abkin2026taskenergy}, whose turn limit matches mini-SWE-agent~\cite{minisweagent}. We retain normal stops, incorrect answers, model failures, and watchdog stops as distinct outcomes, counting success only when independent verification passes.

\begin{table}[t]
\centering\small
\caption{Main live comparisons. Each listed admission limit has the stated number of paired repetitions. Every cohort offers 16 tasks.}
\label{tab:design}
\vskip 0.1in
\begin{tabularx}{\columnwidth}{lCC}
\toprule
\rowcolor{tableheadergray}
& \bfseries Admission & \bfseries Paired\\
\rowcolor{tableheadergray}
\bfseries Precision & \bfseries limit $\boldsymbol{N}$ & \bfseries repetitions\\
\midrule
\multicolumn{3}{@{}l}{\textbf{Devstral Small 1.1, A100 80 GB}}\\[2pt]
BF16 & 4, 16 & 6\\
W8A8 & 4, 8, 16 & 6\\
W4A16 & 2, 4, 8, 16 & 6\\
\addlinespace[4pt]
\multicolumn{3}{@{}l}{\textbf{Devstral Small 1.1, A100 40 GB}}\\[2pt]
W8A8 & 4, 16 & 7\\
W4A16 & 4, 16 & 7\\
\addlinespace[4pt]
\multicolumn{3}{@{}l}{\textbf{Qwen3.5-9B, A100 40 GB}}\\[2pt]
BF16, W8A8, W4A16 & 4 & 11\\
W8A8, W4A16 & 16 & 4\\
\bottomrule
\end{tabularx}
\end{table}

\paragraph{Pairing and uncertainty.}
Within each repetition, we use the same 16 tasks, sampling seed, and admission order across settings, so comparisons hold these inputs fixed. Across repetitions, the seed changes, the task order is reshuffled, and the precision order is balanced. The main 80 GB experiment runs nine settings in each of six repetitions, for 54 cohorts. The admission limits in Table~\ref{tab:design} cover each precision's lowest-cost choices in the pilot runs and all settings selected by the two token-based rules. We run two single-agent cohorts per precision at $N=1$ on 80 GB to test whether sharing the GPU changes the ordering measured with one agent. All limits were fixed before the main runs.

We pool energy and successes as in Equation~\ref{eq:cost}. We obtain 95\% percentile intervals by resampling whole paired repetitions 10,000 times, keeping their tasks and telemetry together. The intervals describe run-to-run variation on the evaluated catalogue.

Before the 40 GB runs, we fixed eight paired repetitions and a rule to include every complete pair. An infrastructure failure left one paired repetition incomplete, leaving seven complete pairs for analysis.

\paragraph{Completion and latency (RQ4).}
We evaluate the selection rules against completion and latency requirements. The prespecified (i.e., fixed in advance) completion reference is BF16 with four admitted agents. The observed completion requirement allows a pooled completion rate at most 6.25 percentage points below this reference, one task per 16-task cohort. A stricter statistical criterion, reported but not required, asks the lower bound of the paired interval for the completion-rate difference to be at least $-6.25$ percentage points.

The prespecified latency requirement is 1,800 seconds for p90 service time, the 90th percentile of elapsed time from admission to session end. Session end includes watchdog cancellation and cleanup, so the rule fails a setting when the watchdog stops at least a tenth of its sessions. A setting is eligible when it meets both the observed completion requirement and this latency requirement. We use these prespecified requirements to evaluate selection rules on the 80 GB Devstral grid. The 40 GB comparisons for Devstral and Qwen3.5-9B examine energy per verified completion across precisions. Queue-inclusive time starts when the cohort's tasks are offered and also counts waiting for admission (Supplementary Section C, Table S5).

For \textbf{RQ4}, we fix two selection rules before the main runs. One chooses the precision with the lowest pilot energy per output token with one agent and serves it at the largest tested admission limit. The other chooses the setting with the highest pilot token throughput. We evaluate these choices against both service requirements and compare their energy per verified completion with the cheapest eligible setting observed on the same data (Section~\ref{sec:selection}).

\paragraph{Replay with fixed recorded work (RQ2).}
An agent can generate different calls at different precisions, so live runs alone cannot isolate the energy of executing the same work. Crossed replay fixes the recorded calls and changes the precision that executes them, called the executing precision. We record each precision's $N=4$ cohort in the first repetition and execute each of these three recorded workloads on all three precisions. Replay reconstructs the exact prompt token IDs, forces the recorded output token IDs, and preserves the order of each session's calls. It replaces tool work with waits lasting the recorded tool and agent-framework time, and its energy includes those waits and server drain. Each of the nine recording/execution combinations is replayed once and reproduces every recorded output-token sequence. Repeated live cohorts provide the uncertainty in task outcomes.

\paragraph{Interventions on cache capacity (RQ3).}
Because W8A8 preempted requests in every primary 40 GB cohort, we ran four further repetitions that add W4A16 with a cache cap at $N=4$ and $N=16$. This exploratory test was outside the original rules, and we compare the cap only within those four repetitions. Restricting W4A16 to W8A8's cache capacity while retaining its model and execution kernels tests whether freed memory contributes to completed work. On 80 GB, an equal-cache comparison gives both quantized precisions BF16's capacity of 176,368 tokens at $N=4$ and $N=16$ over four paired repetitions (Supplementary Section B, Table S4).

\paragraph{Second model family.}
Because the replay and live comparisons could depend on the model family, we also measure Qwen3.5-9B, which differs in family, size, and architecture and has matched published BF16, W8A8, and W4A16 checkpoints~\cite{qwen2026qwen35,redhat2026qwen35w8a8,redhat2026qwen35w4a16}.\footnote{The Qwen team at Alibaba developed Qwen3.5-9B. Red Hat AI supplies the quantized Qwen3.5-9B checkpoints.} Three of every four layers use linear attention, whose per-request state has a fixed size, and the quantized checkpoints keep those layers in 16-bit precision. Its runs use vLLM 0.18.1, the version recorded in its quantization recipes, because vLLM 0.10.2 does not support this architecture. All precisions share the BF16 tokenizer and chat template with thinking disabled, a 65,536-token context limit, and the model card's non-thinking sampling settings (Supplementary Section A). Tool calls use Qwen's own format, and malformed calls receive the same error feedback across its three precisions.

Qwen3.5-9B runs on the 40 GB deployment, which holds all three of its precisions and was available when the 80 GB node was not. Its BF16 provides the completion reference. We retain the same 16 tasks, tools, verification, and stopping rules. Each paired repetition at four admitted agents runs the three precisions live and replays its W8A8 and W4A16 recordings on both quantized precisions. Repeating the live runs measures variation between recordings. For two paired repetitions, we execute the recordings again to measure variation between executions. For replay comparisons, duplicate executions are averaged. We pool the 11 completed repetitions. In four completed paired repetitions, we compare W8A8 and W4A16 with 16 admitted agents to test whether greater admission creates cache pressure.

\paragraph{Energy measurement and inclusion checks.}
We measure gross energy from the CPU-package counters and the serving GPU's board-energy counter at a nominal 10 Hz, alongside server telemetry at 1 Hz. The interval defined in Section~\ref{sec:metric} includes tool work and idle consumption and excludes model loading and independent verification. Mean cohort power varies little within a deployment, 430 to 447 W across the 80 GB cohorts, so energy differences largely reflect elapsed time. With eight or more admitted agents, 29 of the 30 cohorts on 80 GB last until the 1,800-second watchdog stops their last session. Before retaining a cohort, we check that both energy sources cover the interval, the server drained, all 16 tasks are accounted for, and cleanup completed. Supplementary Section B reports the supporting checks.

\section{Results}
\label{sec:results}
Energy and cost ratios compare W4A16 with W8A8, so values below one favor W4A16.
\subsection{\texorpdfstring{RQ1: Whether the Cheapest Precision Depends on Admission and Deployment}{RQ1: Whether the Cheapest Precision Depends on Admission and Deployment}}
\label{sec:outcomes}
\begin{figure}[t]
\centering
\includegraphics[width=\columnwidth,trim=0 4.1bp 0 2.8bp,clip]{\paperpath 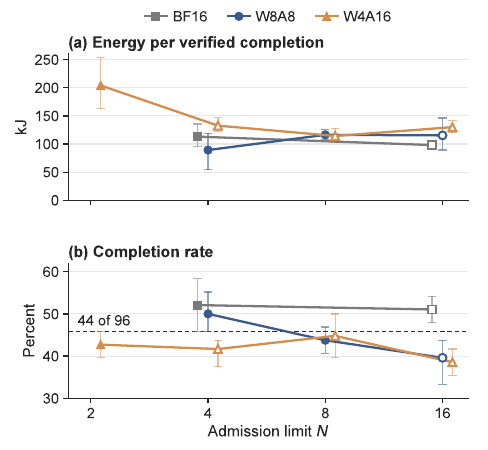}
\caption{On the 80 GB deployment, the lowest observed cost occurs at $N=4$ for W8A8, $N=8$ for W4A16, and $N=16$ for BF16. (a) Energy per verified completion and (b) completion rate across six paired repetitions with 95\% intervals. Marks are offset horizontally for legibility. The dashed line marks the minimum completion requirement of 44 verified completions out of 96 offered task attempts. Open marks indicate configurations that fail the 1,800-second latency requirement. Only filled marks on or above the dashed line satisfy both requirements.}
\label{fig:admission}
\end{figure}

On 80 GB, the precision with the lowest observed cost changes from W8A8 with four admitted agents to BF16 with 16 admitted agents (Figure~\ref{fig:admission}). Each precision reaches its minimum at a different admission limit, $N=4$ for W8A8, $N=8$ for W4A16, and $N=16$ for BF16. Resampling the six paired repetitions reproduces all three choices together in 77\% of resamples. All three precisions share a lowest-cost admission limit in only 0.2\% of resamples. Beyond four agents, most cohorts run until the watchdog, so cost depends mainly on the number of verified completions within the 1,800-second limit. Where completion counts remain relatively stable, increasing admission can reduce cost, which occurs for BF16 between $N=4$ and $N=16$ and for W4A16 between $N=2$ and $N=8$. 
Among the tested 80 GB configurations, W8A8 at $N=4$ achieves the lowest cost of 89.2 kJ per verified completion, with 48 of 96 task attempts passing verification. At $N=8$ and $N=16$, its success count falls to 42 and 38 verified completions, respectively.

\begin{figure}[t]
\centering
\includegraphics[width=\columnwidth,trim=0 4.7bp 0 6.3bp,clip]{\paperpath 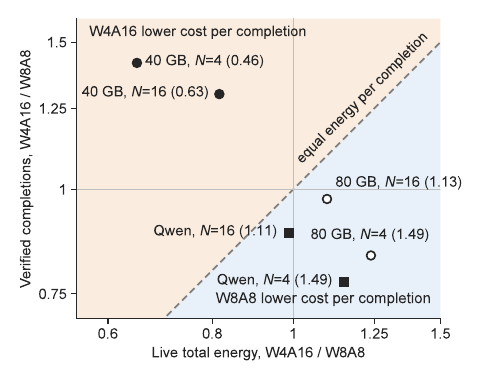}
\caption{With four agents where requests fit in the cache, W4A16's higher live energy and lower completion count raise its cost. Points give W4A16/W8A8 ratios of completions against live energy, with the ratio of energy per verified completion beside each point. Circles are Devstral, filled for 40 GB and open for 80 GB, and squares are Qwen3.5-9B on 40 GB. Above the dashed diagonal, W4A16 costs less.}
\label{fig:boundary}
\end{figure}

\begin{figure}[!tb]
\centering
\includegraphics[width=\columnwidth,trim=0 8.2bp 0 0.0bp,clip]{\paperpath 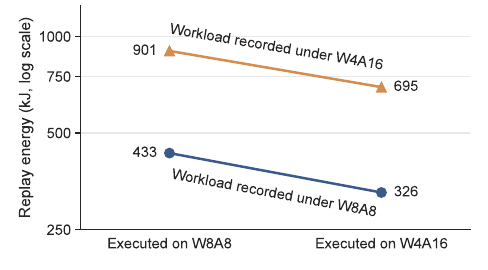}
\caption{The Devstral workload recorded under W4A16 in the first repetition uses about twice the replay energy of the one recorded under W8A8 on either executing precision. Each line holds one recorded workload fixed and changes the executing precision (full matrix in Supplementary Section B, Table S3).}
\label{fig:replay}
\end{figure}

\begin{figure*}[t]
\centering
\includegraphics[width=\textwidth,trim=0 5.5bp 0 0.0bp,clip]{\paperpath 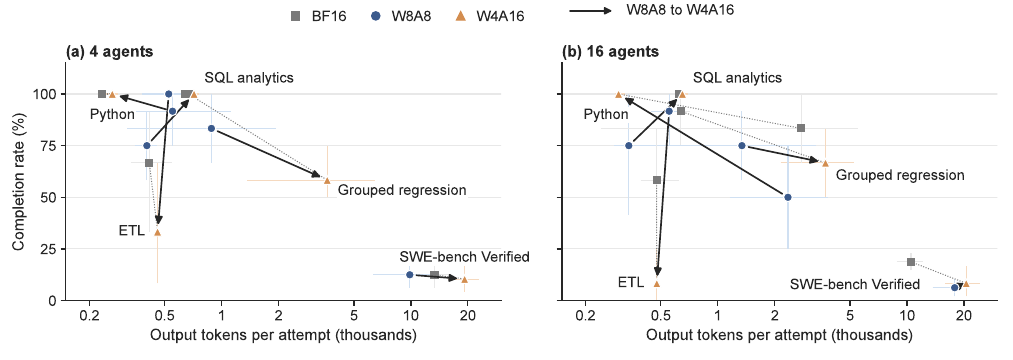}\\[2pt]
\hspace*{23bp}\includegraphics[width=0.9396\textwidth,trim=0 0 29.38bp 0,clip]{\paperpath 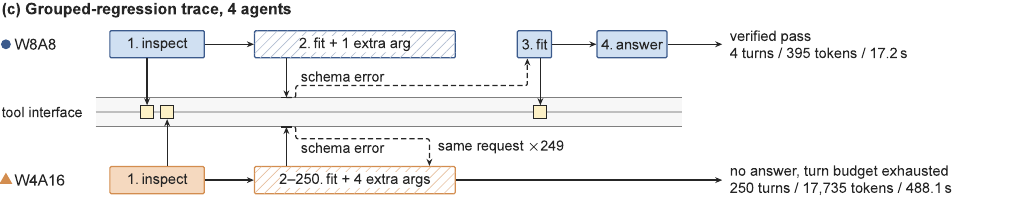}%
\caption{Task outcomes across precisions and contrasting responses to a tool error on 80 GB. (a, b) Completion rate against output tokens per attempt, with arrows from W8A8 to W4A16, gray squares for BF16 tied by dotted lines to each family's W4A16 mark, and 95\% intervals over six repetitions. (c) A pair of grouped-regression attempts with four admitted agents, selected after the runs, shows correction versus repetition after a schema error. W8A8 corrects its call and passes verification, while W4A16 repeats its invalid call until the turn limit. Inspect reads dataset metadata, and fit requests a regression. Hatched boxes mark invalid tool requests that fail validation, dashed arrows return schema errors, and squares on the tool lane mark executed calls. Numbered boxes are agent turns, read from left to right, and W4A16's repeated requests share one box.}
\label{fig:behavior}
\end{figure*}

\noindent\textbf{Single-agent cohorts.} Across two single-agent cohorts per precision on 80 GB, every precision costs more per verified completion than in its tested shared settings. W4A16 costs less than BF16 with one agent, while BF16 costs less at their lowest observed shared settings (Supplementary Section A).

\noindent\textbf{W4A16 on 80 GB.} W4A16 costs the most of the three precisions at $N=4$ and $N=16$ on 80 GB. At $N=16$, nearly every cohort reaches the watchdog. Over six cohorts, W4A16 consumes nearly the same total energy as BF16 but completes 12 fewer tasks (Figure~\ref{fig:admission}). It costs 31.9 kJ more per verified completion (paired 95\% interval [19.3, 45.5] kJ). At $N=4$, W4A16 costs 43.4 kJ more per verified completion than W8A8 (paired interval [14.5, 75.8] kJ). At $N=16$, their cost difference remains unresolved because its interval includes zero. We trace why the four-agent live comparison favors W8A8 despite W4A16's saving on fixed calls (Section~\ref{sec:work}).

\noindent\textbf{The 40 GB deployment.} On the 40 GB deployment, Devstral's W4A16 configuration costs less than W8A8 at both tested admission limits. Across the seven primary paired repetitions, W4A16 reduces energy per completion by 54\% at $N=4$ and 37\% at $N=16$, with paired ratio intervals of [0.30, 0.67] and [0.43, 0.82], respectively. W4A16 completes more tasks at both limits. We test whether the additional cache contributes to this advantage (Section~\ref{sec:capacity}).

\takeaway{1}{The cheapest precision depends on admission and deployment. On 80 GB, W8A8 at $N=4$ has the lowest observed cost overall (Figure~\ref{fig:admission}). On 40 GB, Devstral's W4A16 costs less than W8A8 at both tested limits (Figure~\ref{fig:preview}). Operators should reassess the choice of precision when changing the admission limit or GPU configuration.}

\subsection{\texorpdfstring{RQ2: Whether Fixed-Call Savings Reduce Energy per Verified Completion}{RQ2: Whether Fixed-Call Savings Reduce Energy per Verified Completion}}
\label{sec:work}
With four admitted agents on 80 GB, W4A16 executes recorded calls with less energy than W8A8 but costs more per verified completion in live runs (Figures~\ref{fig:preview} and~\ref{fig:boundary}). W4A16 consumes 23.9\% more total energy than W8A8 and completes 16.7\% fewer tasks. These two changes raise energy per completion by 48.6\% (Equation~\ref{eq:boundary}).

\noindent\textbf{Fixed recorded calls.} To explain the higher live cost, we compare precisions on the same recorded calls (Figure~\ref{fig:replay}). The recorded work changes replay energy more than the executing precision does. W4A16 uses less energy than W8A8 for all three recorded workloads: 24.6\% less for each workload recorded under BF16 and W8A8, and 22.9\% less for the workload recorded under W4A16 (Figure~\ref{fig:preview}). On either executing precision, the workload recorded under W4A16 in the first repetition uses about twice the replay energy of the workload recorded under W8A8 (Figure~\ref{fig:replay}). In these replays, that difference outweighs the saving from executing on W4A16. Each workload comes from a single live cohort, so replay compares execution of those recordings while the six live repetitions measure variation in task outcomes.

\noindent\textbf{Generated work.} Across six paired repetitions at $N=4$, W4A16's agents generate 0.98 million output tokens compared to 0.50 million output tokens for W8A8 (Figure~\ref{fig:behavior}a). Repository tasks account for most of these tokens but yield only five verified completions for W4A16 and six verified completions for W8A8.

\noindent\textbf{Task families.} At $N=4$, W4A16 gains completions on Python and SQL analytics and loses them on ETL and grouped regression (Figure~\ref{fig:behavior}a). It completes all Python and SQL-analytics attempts, but only four of 12 ETL attempts, compared to all 12 ETL attempts for W8A8. This eight-task ETL gap matches W4A16's overall gap in verified completions. At $N=16$, W4A16 still completes all Python and SQL-analytics attempts, while the ETL gap widens to just one completion for W4A16 against 11 completions for W8A8.

\noindent\textbf{A recovery trace.} Figure~\ref{fig:behavior}c shows how different responses to the same tool error can lead to completion or repeated failure. Both precisions inspect metadata for the million-row dataset and then request a regression fit. The schema rejects extra arguments before executing the regression. W8A8 removes its extra argument, executes the fit, and returns a verified answer in four turns and 395 tokens. W4A16 repeats its invalid request until it exhausts the 250-turn budget, generating 17,735 tokens. Its 249 rejected proposals execute no regression, and only the initial metadata inspection completes. Across all six repetitions of this task, W8A8 passes six times and W4A16 passes once. \citet{jang2026flatscore} find that 4-bit quantization increases tool-call failures for some model and task-domain combinations.

\begin{figure}[t]
\centering
\includegraphics[width=\columnwidth]{\paperpath 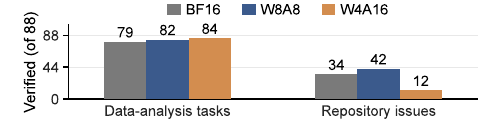}
\caption{For Qwen3.5-9B, W4A16's completion deficit comes from repository issues. Bars pool 11 paired repetitions with four admitted agents on 40 GB, with 88 attempts per task type and precision.}
\label{fig:qwen}
\end{figure}

\begin{figure}[tb]
\centering
\includegraphics[width=\columnwidth,trim=0 2.5bp 0 0.0bp,clip]{\paperpath 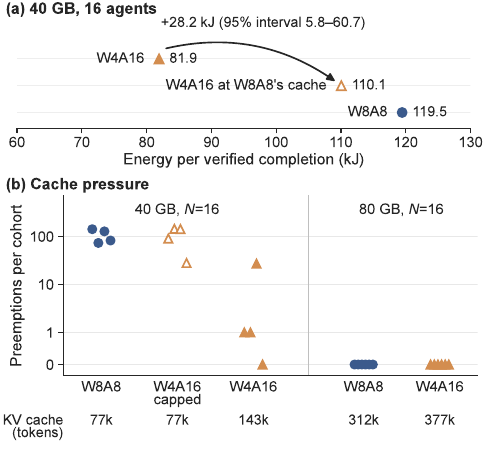}
\caption{Capping W4A16's cache raises its energy per verified completion with 16 agents on 40 GB. (a) Pooled over the four repetitions that ran all three settings, with the cap's paired effect and 95\% interval. (b) Preemptions per cohort at 16 agents on a symmetric-log axis, with each group's cache capacity below it.}
\label{fig:capacity}
\end{figure}

\noindent\textbf{A second model.} We repeat live runs and replay with Qwen3.5-9B to test whether its generated work offsets savings on recorded calls. Across 22 Qwen3.5-9B workloads recorded in 11 paired repetitions at $N=4$ on the 40 GB deployment, W4A16 executes the same recorded calls with 9.8--12.3\% less energy than W8A8 (Figure~\ref{fig:preview}). On either executing precision, the workloads recorded under W4A16 use on average 1.37 times the replay energy of those recorded under W8A8.

In Qwen3.5-9B's live cohorts, W4A16 costs 49\% more per verified completion than W8A8 (95\% ratio interval $[1.28,1.72]$). It generates 30\% more output tokens and uses 15\% more total energy. W4A16 completes 96 of 176 task attempts, while W8A8 completes 124 task attempts. The deficit comes from repository issues, where W4A16 completes 30 fewer task attempts than W8A8 (Figure~\ref{fig:qwen}). Devstral's overall deficit comes from data-analysis tasks. BF16 completes 113 task attempts and costs 41.3 kJ per verified completion. W8A8 completes more tasks than BF16 at a lower cost, 32.6 kJ per verified completion, making it the cheapest precision in this comparison. All 33 Qwen3.5-9B cohorts run without preemption, so their requests fit in the cache.

\takeaway{2}{Generated work and lost completions can outweigh cheaper execution. In our comparisons with four admitted agents where requests fit in the cache, W4A16 executes the same calls with 10--25\% less energy than W8A8 but costs 49\% more per verified completion in live runs, with paired 95\% intervals favoring W8A8 (Figures~\ref{fig:preview}, \ref{fig:boundary}, and~\ref{fig:qwen}). An operator should pair replay measurements with live outcomes before choosing a precision.}

\subsection{\texorpdfstring{RQ3: When Freed Memory Lowers the Cost}{RQ3: When Freed Memory Lowers the Cost}}
\label{sec:capacity}

On 40 GB, Devstral's W4A16 costs less than W8A8. Capping its cache shows that freed memory contributes to this advantage at 16 agents (Figure~\ref{fig:capacity}). In four later repetitions of all three settings, the cap raises W4A16's pooled energy per verified completion from 81.9 to 110.1 kJ. The increase is 28.2 kJ per verified completion (34.4\%), with a paired interval of $[5.8,60.7]$ kJ. W8A8 uses 119.5 kJ per verified completion. Cost increases in three of the four repetitions.

\noindent\textbf{Cache pressure.} W8A8 and capped W4A16 each have a 77,056-token cache, which holds little more than one full 65,536-token context, and W8A8 preempts requests in every 40 GB cohort at both admission limits. With so little cache, they have 28 to 143 preemptions per cohort at 16 agents, against 0 to 27 preemptions per cohort for uncapped W4A16 (Figure~\ref{fig:capacity}b). These four repetitions leave two comparisons unresolved, capped W4A16 against W8A8 and the effect of the cap at four admitted agents, whose sign varies across repetitions.

\noindent\textbf{Qwen3.5-9B at 16 agents.} On the same 40 GB GPU, Qwen3.5-9B's smaller weights and fixed-size linear-attention state give even its W8A8 a 189,552-token cache, against 77,056 tokens for Devstral's W8A8 and 142,608 tokens for Devstral's W4A16. With 16 admitted agents, its requests still fit. Across four paired repetitions, every cohort runs without preemption, and cache use peaks at 41\%. W4A16 uses 0.99 times W8A8's energy and completes 39 of 64 task attempts, while W8A8 completes 44 task attempts. W4A16 costs 1.11 times as much per verified completion as W8A8, with a 95\% interval of $[0.93,1.49]$, so the ordering is unresolved. Compared with separate four-agent repetitions, energy per verified completion falls from 32.6 to 24.0 kJ for W8A8 and from 48.4 to 26.8 kJ for W4A16.

\noindent\textbf{Normal and capped cache on 80 GB.} With both quantized precisions capped at BF16's capacity, their ordering is unresolved (Supplementary Section B, Table S4). At normal cache capacity, the 80 GB quantized cohorts show little pressure. All 42 of these cohorts run without preemption, and at most 0.72\% of any cohort's active telemetry samples have a waiting request.

\takeaway{3}{Freed memory can lower the cost where requests compete for cache. On 40 GB with 16 Devstral agents, where W8A8 preempts requests in every cohort, capping W4A16's cache at W8A8's capacity raises its pooled energy per verified completion by 34\% over four repetitions (Figure~\ref{fig:capacity}). Extra capacity has little cache pressure to relieve where requests already fit, as they do on 80 GB and for Qwen3.5-9B even with 16 agents.}

\subsection{\texorpdfstring{RQ4: What Choosing by Token Measurements Misses}{RQ4: What Choosing by Token Measurements Misses}}
\label{sec:selection}
\begin{table}[!htbp]
\centering\footnotesize
\caption{Among these 80 GB settings, only BF16 and W8A8 at $N=4$ meet the 1,800-second latency requirement. W8A8 at $N=4$ (bold) is cheapest among settings meeting both service requirements. $\dagger$ marks the per-token choice, and $\ddagger$ marks the throughput choice. Verified completions are out of 96 task attempts. Energy includes attempts stopped by the watchdog.}
\label{tab:decision}
\vskip 0.1in
\setlength{\tabcolsep}{0.85pt}
\begin{tabularx}{\columnwidth}{@{}l>{\hsize=.75\hsize\raggedleft\arraybackslash}X>{\hsize=1.15\hsize\raggedleft\arraybackslash}X>{\hsize=1.15\hsize\raggedleft\arraybackslash}X>{\hsize=.95\hsize\raggedleft\arraybackslash}X@{}}
\toprule
\rowcolor{tableheadergray}
& \centering\arraybackslash\bfseries Verified & \centering\arraybackslash\bfseries kJ per & \centering\arraybackslash\bfseries p90 & \centering\arraybackslash\bfseries Watchdog\\
\rowcolor{tableheadergray}
\bfseries Setting & \centering\arraybackslash\bfseries of 96 & \centering\arraybackslash\bfseries completion & \centering\arraybackslash\bfseries service (s) & \centering\arraybackslash\bfseries stops\\
\midrule
BF16, $N=4$ & 50 & 113.1 & 1,594 & 9\\
BF16, $N=16$ & 49 & 98.1 & 1,802 & 28\\
\textbf{W8A8, $N=4$} & \textbf{48} & \textbf{89.2} & \textbf{858} & \textbf{7}\\
W8A8, $N=16$$^{\ddagger}$ & 38 & 115.3 & 1,802 & 10\\
W4A16, $N=4$ & 40 & 132.6 & 1,802 & 11\\
W4A16, $N=16$$^{\dagger}$ & 37 & 130.0 & 1,802 & 16\\
\bottomrule
\end{tabularx}
\end{table}

On 80 GB, both prespecified rules miss the required 44 verified completions out of 96 task attempts and fail the latency requirement (Table~\ref{tab:decision}). The rule based on energy per output token selects W4A16 at $N=16$. Its completion rate is 13.5 percentage points below BF16 (interval $[-19.8,-5.2]$ percentage points). The throughput rule selects W8A8 at $N=16$, whose completion rate is 12.5 percentage points below BF16 (interval $[-21.9,-3.1]$ percentage points). Compared with W8A8 at $N=4$, the cheapest eligible setting selected on these same data, the per-token choice costs 45.6\% more per verified completion (ratio interval $[1.13,2.29]$). The throughput choice's observed 29.2\% higher cost is unresolved (ratio interval $[0.82,2.32]$).

\noindent\textbf{The eligible setting.} W8A8 at $N=4$ completes two fewer tasks than BF16 and meets the observed completion and latency requirements (Table~\ref{tab:decision}). Eligibility uses the observed completion rate, while the stricter statistical criterion accounts for uncertainty. Its completion-rate interval, $[-10.4,6.3]$ percentage points relative to BF16, extends below the stricter $-6.25$-percentage-point threshold, so that criterion is not met. Its energy difference from BF16 also remains unresolved because the interval includes zero ($[-74.7,13.8]$ kJ per verified completion).

\takeaway{4}{Both rules select settings that fail the completion and latency requirements on 80 GB. The per-token choice also costs 45.6\% more per verified completion than the cheapest eligible setting observed on the same data (Table~\ref{tab:decision}). Operators should identify settings that meet their service's completion and latency requirements before comparing their energy per verified completion.}

\section{Discussion} %
\label{sec:discussion}
A setting's energy per verified completion reflects its mix of tasks. The breakdown by task family helps an operator judge whether a lower aggregate cost preserves the outcomes the service needs (Section~\ref{sec:work}). A repository-repair service, for example, needs to assess repair completions. Completion requirements can reflect these application needs before energy is compared among eligible settings (Section~\ref{sec:selection}).

Memory freed by quantization can also lower energy costs by reducing preemption and recovery work when admitted requests compete for the KV cache (Section~\ref{sec:capacity}). Where requests already fit, extra capacity has little cache pressure to relieve, and differences in generated work and successful completions can outweigh cheaper execution. Deployment comparisons should report cache pressure alongside memory capacity, because the model, context lengths, and admission limit determine how that capacity is used.

Our ratios compare precisions within a model and deployment. Comparisons between models are descriptive because their engine versions, sampling settings, and tool-call formats also differ. The experiments use a 16-task catalog offered in cohorts. A service with continuing arrivals should evaluate its expected arrival pattern, which can change concurrency and cache pressure. Pairing replay with verified live runs also offers a way to examine other serving optimizations that change the work agents generate.

\section{Conclusion} %
The energy benefit of quantization depends on both the work agents generate and the memory available to concurrent requests. Cheaper execution can be outweighed by extra calls and lost completions, while additional cache capacity can help when requests compete for memory. An operator should compare candidate precisions and admission limits on representative tasks under the service's expected load. Settings should first meet that service's completion and latency requirements. Among those settings, the operator should select the one with the lowest measured energy per verified completion, including energy spent on failed attempts. The comparison should be repeated when the model, workload, or available memory changes.

\FloatBarrier
\bibliographystyle{\paperpath mlsys2025}
\bibliography{\paperpath refs}

\begin{thebibliography}{28}
\providecommand{\natexlab}[1]{#1}
\providecommand{\url}[1]{\texttt{#1}}
\expandafter\ifx\csname urlstyle\endcsname\relax
  \providecommand{\doi}[1]{doi: #1}\else
  \providecommand{\doi}{doi: \begingroup \urlstyle{rm}\Url}\fi

\bibitem[Abkin et~al.(2026)Abkin, Yang, Flesher, Kodarapu, Panigrahy, Tyagi,
  and Houngninou]{abkin2026taskenergy}
Abkin, S., Yang, A., Flesher, A., Kodarapu, N., Panigrahy, D., Tyagi, A., and
  Houngninou, D.~K.
\newblock Measuring the energy cost of completing tasks with quantized
  language-model agents, 2026.
\newblock URL \url{https://arxiv.org/abs/2610.32417}.

\bibitem[Agrawal et~al.(2024)Agrawal, Kedia, Panwar, Mohan, Kwatra, Gulavani,
  Tumanov, and Ramjee]{agrawal2024sarathi}
Agrawal, A., Kedia, N., Panwar, A., Mohan, J., Kwatra, N., Gulavani, B.~S.,
  Tumanov, A., and Ramjee, R.
\newblock Taming {Throughput-Latency} tradeoff in {LLM} inference with
  {Sarathi-Serve}.
\newblock In \emph{18th USENIX Symposium on Operating Systems Design and
  Implementation (OSDI 24)}, pp.\  117--134. USENIX Association, 2024.
\newblock URL
  \url{https://www.usenix.org/conference/osdi24/presentation/agrawal}.

\bibitem[Chowdhury et~al.(2024)Chowdhury, Aung, Chan, Jaffe, Sherburn, Starace,
  Mays, Dias, Aljubeh, Glaese, Jimenez, Yang, Ho, Patwardhan, Liu, and
  Madry]{openai2024swebenchverified}
Chowdhury, N., Aung, J., Chan, J.~S., Jaffe, O., Sherburn, D., Starace, G.,
  Mays, E., Dias, R., Aljubeh, M., Glaese, M., Jimenez, C.~E., Yang, J., Ho,
  L., Patwardhan, T., Liu, K., and Madry, A.
\newblock Introducing {SWE-bench Verified}.
\newblock OpenAI research announcement. Updated February 24, 2025, 2024.
\newblock URL \url{https://openai.com/index/introducing-swe-bench-verified/}.

\bibitem[Chung et~al.(2025)Chung, Ma, Wu, Liu, Kweon, Xia, Wu, and
  Chowdhury]{chung2025mlenergy}
Chung, J.-W., Ma, J.~J., Wu, R., Liu, J., Kweon, O.~J., Xia, Y., Wu, Z., and
  Chowdhury, M.
\newblock The {ML.ENERGY} benchmark: Toward automated inference energy
  measurement and optimization.
\newblock In Belgrave, D., Zhang, C., Lin, H., Pascanu, R., Koniusz, P.,
  Ghassemi, M., and Chen, N. (eds.), \emph{Advances in Neural Information
  Processing Systems}, volume 38, Main Conference. Curran Associates, Inc.,
  2025.
\newblock \doi{10.52202/085713-3652}.
\newblock URL
  \url{https://proceedings.neurips.cc/paper_files/paper/2025/file/9dc510e3d7b0b3b2a58ffed7a3ad6b0f-Paper-Datasets_and_Benchmarks_Track.pdf}.

\bibitem[Chung et~al.(2026)Chung, Wu, Ma, and Chowdhury]{chung2026joules}
Chung, J.-W., Wu, R., Ma, J.~J., and Chowdhury, M.
\newblock Where do the joules go? diagnosing inference energy consumption,
  2026.
\newblock URL \url{https://arxiv.org/abs/2601.22076v2}.
\newblock Version 2.

\bibitem[Frantar et~al.(2023)Frantar, Ashkboos, Hoefler, and
  Alistarh]{frantar2023gptq}
Frantar, E., Ashkboos, S., Hoefler, T., and Alistarh, D.
\newblock {OPTQ}: Accurate quantization for generative pre-trained
  transformers.
\newblock In \emph{The Eleventh International Conference on Learning
  Representations}, 2023.
\newblock URL \url{https://openreview.net/forum?id=tcbBPnfwxS}.

\bibitem[Gao et~al.(2025)Gao, Zhou, Sun, Zhang, and Wen]{gao2025rethinkingkv}
Gao, W., Zhou, X., Sun, P., Zhang, T., and Wen, Y.
\newblock Rethinking key-value cache compression techniques for large language
  model serving.
\newblock In \emph{Proceedings of Machine Learning and Systems}, volume~7.
  MLSys, 2025.
\newblock URL \url{https://arxiv.org/abs/2503.24000}.

\bibitem[Ifath \& Haque(2026)Ifath and Haque]{ifath2026multirequest}
Ifath, M. M.~A. and Haque, I.
\newblock Characterizing performance--energy trade-offs of large language
  models in multi-request workflows.
\newblock \emph{Proceedings of the ACM on Measurement and Analysis of Computing
  Systems}, 10\penalty0 (1):\penalty0 1--26, 2026.
\newblock \doi{10.1145/3788089}.
\newblock Article 7.

\bibitem[Jang et~al.(2026)Jang, Yang, Lim, and Park]{jang2026flatscore}
Jang, J., Yang, K., Lim, H., and Park, H.
\newblock Flat score, amplified failures: How the error budget masks damage in
  quantized {LLM} agents, 2026.
\newblock URL \url{https://arxiv.org/abs/2607.27275}.

\bibitem[Jimenez et~al.(2024)Jimenez, Yang, Wettig, Yao, Pei, Press, and
  Narasimhan]{jimenez2024swebench}
Jimenez, C.~E., Yang, J., Wettig, A., Yao, S., Pei, K., Press, O., and
  Narasimhan, K.
\newblock {SWE-bench}: Can language models resolve real-world {GitHub} issues?
\newblock In Kim, B., Yue, Y., Chaudhuri, S., Fragkiadaki, K., Khan, M., and
  Sun, Y. (eds.), \emph{International Conference on Learning Representations},
  volume 2024, pp.\  54107--54157, 2024.
\newblock URL
  \url{https://proceedings.iclr.cc/paper_files/paper/2024/file/edac78c3e300629acfe6cbe9ca88fb84-Paper-Conference.pdf}.

\bibitem[Kapoor et~al.(2025)Kapoor, Stroebl, Siegel, Nadgir, and
  Narayanan]{kapoor2025agents}
Kapoor, S., Stroebl, B., Siegel, Z.~S., Nadgir, N., and Narayanan, A.
\newblock {AI} agents that matter.
\newblock \emph{Transactions on Machine Learning Research}, 2025.
\newblock ISSN 2835-8856.
\newblock URL \url{https://openreview.net/forum?id=Zy4uFzMviZ}.

\bibitem[Kwon et~al.(2023)Kwon, Li, Zhuang, Sheng, Zheng, Yu, Gonzalez, Zhang,
  and Stoica]{kwon2023pagedattention}
Kwon, W., Li, Z., Zhuang, S., Sheng, Y., Zheng, L., Yu, C.~H., Gonzalez, J.~E.,
  Zhang, H., and Stoica, I.
\newblock Efficient memory management for large language model serving with
  {PagedAttention}.
\newblock In \emph{Proceedings of the 29th Symposium on Operating Systems
  Principles}, pp.\  611--626, 2023.
\newblock \doi{10.1145/3600006.3613165}.

\bibitem[Lin et~al.(2024)Lin, Han, Zhang, Yang, Yang, Chen, and
  Qiu]{lin2024parrot}
Lin, C., Han, Z., Zhang, C., Yang, Y., Yang, F., Chen, C., and Qiu, L.
\newblock Parrot: Efficient serving of {LLM-based} applications with semantic
  variable.
\newblock In \emph{18th USENIX Symposium on Operating Systems Design and
  Implementation (OSDI 24)}, pp.\  929--945. USENIX Association, 2024.
\newblock URL
  \url{https://www.usenix.org/conference/osdi24/presentation/lin-chaofan}.

\bibitem[Lin et~al.(2025)Lin, Tang, Yang, Zhang, Xiao, Gan, and
  Han]{MLSYS2025_fbe2b2f7}
Lin, Y., Tang, H., Yang, S., Zhang, Z., Xiao, G., Gan, C., and Han, S.
\newblock {QServe}: {W4A8KV4} quantization and system co-design for efficient
  {LLM} serving.
\newblock In Zaharia, M., Joshi, G., and Lin, Y. (eds.), \emph{Proceedings of
  Machine Learning and Systems}, volume~7. MLSys, 2025.
\newblock URL
  \url{https://proceedings.mlsys.org/paper_files/paper/2025/file/fbe2b2f74a2ece8070d8fb073717bda6-Paper-Conference.pdf}.

\bibitem[Liu et~al.(2026)Liu, Yu, Park, Stoica, and Cheung]{MLSYS2026_554e056f}
Liu, X., Yu, J., Park, J., Stoica, I., and Cheung, A.
\newblock Speculative decoding: Performance or illusion?
\newblock In Chowdhery, A. and Jia, Z. (eds.), \emph{Proceedings of Machine
  Learning and Systems}, volume~8, pp.\  1719--1741. MLSys, 2026.
\newblock URL
  \url{https://proceedings.mlsys.org/paper_files/paper/2026/file/554e056fe2b6d9fd27ffcd3367ae1267-Paper-Conference.pdf}.

\bibitem[{Mistral AI}(2025)]{mistral2025devstralsmall2}
{Mistral AI}.
\newblock {Devstral Small 2 24B Instruct 2512}: Model card.
\newblock Hugging Face model repository, 2025.
\newblock URL
  \url{https://huggingface.co/mistralai/Devstral-Small-2-24B-Instruct-2512}.

\bibitem[{Mistral AI} \& {All Hands AI}(2025){Mistral AI} and {All Hands
  AI}]{mistral2025devstral11}
{Mistral AI} and {All Hands AI}.
\newblock {Devstral Small 1.1}: Model card.
\newblock Hugging Face model repository, 2025.
\newblock URL
  \url{https://huggingface.co/mistralai/Devstral-Small-2507/tree/bd165ab26cebbcc2eea2c4ecbfc07f3ac42b3c39}.

\bibitem[{Qwen Team}(2026)]{qwen2026qwen35}
{Qwen Team}.
\newblock {Qwen3.5-9B}: Model card.
\newblock Hugging Face model repository, 2026.
\newblock URL
  \url{https://huggingface.co/Qwen/Qwen3.5-9B/tree/c202236235762e1c871ad0ccb60c8ee5ba337b9a}.

\bibitem[{Red Hat AI}(2025{\natexlab{a}})]{redhat2025devstralw4a16}
{Red Hat AI}.
\newblock {Devstral-Small-2507-quantized.w4a16}: Model card and quantization
  recipe.
\newblock Hugging Face model repository, 2025{\natexlab{a}}.
\newblock URL
  \url{https://huggingface.co/RedHatAI/Devstral-Small-2507-quantized.w4a16/tree/bb96f0d1a876d6f6f3ed657b6a965328697ba2f9}.

\bibitem[{Red Hat AI}(2025{\natexlab{b}})]{redhat2025devstralw8a8}
{Red Hat AI}.
\newblock {Devstral-Small-2507-quantized.w8a8}: Model card and quantization
  recipe.
\newblock Hugging Face model repository, 2025{\natexlab{b}}.
\newblock URL
  \url{https://huggingface.co/RedHatAI/Devstral-Small-2507-quantized.w8a8/tree/b152b0574658cdd159721472ebf5353d8d199451}.

\bibitem[{Red Hat AI}(2026{\natexlab{a}})]{redhat2026qwen35w4a16}
{Red Hat AI}.
\newblock {Qwen3.5-9B-quantized.w4a16}: Model card and quantization recipe.
\newblock Hugging Face model repository, 2026{\natexlab{a}}.
\newblock URL
  \url{https://huggingface.co/RedHatAI/Qwen3.5-9B-quantized.w4a16/tree/a398088c4228b0ae0c8c78df88fd1e4bf445f068}.

\bibitem[{Red Hat AI}(2026{\natexlab{b}})]{redhat2026qwen35w8a8}
{Red Hat AI}.
\newblock {Qwen3.5-9B-quantized.w8a8}: Model card and quantization recipe.
\newblock Hugging Face model repository, 2026{\natexlab{b}}.
\newblock URL
  \url{https://huggingface.co/RedHatAI/Qwen3.5-9B-quantized.w8a8/tree/4257ccf0e5c6f87246fd96ea2e4fb16817101558}.

\bibitem[Siraj \& Barbati(2026)Siraj and Barbati]{siraj2026smaller}
Siraj, S. and Barbati, M.
\newblock When is smaller greener? right-sizing language models for agentic
  {AI}, 2026.
\newblock URL \url{https://arxiv.org/abs/2610.19345}.
\newblock Preprint.

\bibitem[Su et~al.(2026)Su, Zhang, Lan, Wang, Shen, Yang, and
  Cheng]{MLSYS2026_8144a9d6}
Su, Z., Zhang, Z., Lan, T., Wang, Z., Shen, H., Yang, J., and Cheng, Y.
\newblock {MorphServe}: Efficient and workload-aware {LLM} serving via runtime
  quantized layer swapping and {KV} cache resizing.
\newblock In Chowdhery, A. and Jia, Z. (eds.), \emph{Proceedings of Machine
  Learning and Systems}, volume~8, pp.\  1056--1075. MLSys, 2026.
\newblock URL
  \url{https://proceedings.mlsys.org/paper_files/paper/2026/file/8144a9d62e506af0fcdeac0e456b2710-Paper-Conference.pdf}.

\bibitem[{SWE-agent contributors}(2025)]{minisweagent}
{SWE-agent contributors}.
\newblock mini-{SWE}-agent.
\newblock GitHub repository, 2025.
\newblock URL \url{https://github.com/SWE-agent/mini-swe-agent}.

\bibitem[Xiao et~al.(2023)Xiao, Lin, Seznec, Wu, Demouth, and
  Han]{xiao2023smoothquant}
Xiao, G., Lin, J., Seznec, M., Wu, H., Demouth, J., and Han, S.
\newblock {SmoothQuant}: Accurate and efficient post-training quantization for
  large language models.
\newblock In \emph{Proceedings of the 40th International Conference on Machine
  Learning}, volume 202 of \emph{Proceedings of Machine Learning Research},
  pp.\  38087--38099. PMLR, 2023.
\newblock URL \url{https://proceedings.mlr.press/v202/xiao23c.html}.

\bibitem[Yang et~al.(2024)Yang, Jimenez, Wettig, Lieret, Yao, Narasimhan, and
  Press]{yang2024sweagent}
Yang, J., Jimenez, C.~E., Wettig, A., Lieret, K., Yao, S., Narasimhan, K., and
  Press, O.
\newblock {SWE-agent}: Agent-computer interfaces enable automated software
  engineering.
\newblock In \emph{Advances in Neural Information Processing Systems},
  volume~37, pp.\  50528--50652. Curran Associates, Inc., 2024.
\newblock \doi{10.52202/079017-1601}.
\newblock URL
  \url{https://proceedings.neurips.cc/paper_files/paper/2024/hash/5a7c947568c1b1328ccc5230172e1e7c-Abstract-Conference.html}.

\bibitem[Yu et~al.(2022)Yu, Jeong, Kim, Kim, and Chun]{yu2022orca}
Yu, G.-I., Jeong, J.~S., Kim, G.-W., Kim, S., and Chun, B.-G.
\newblock Orca: A distributed serving system for {Transformer-Based} generative
  models.
\newblock In \emph{16th USENIX Symposium on Operating Systems Design and
  Implementation (OSDI 22)}, pp.\  521--538. USENIX Association, 2022.
\newblock URL \url{https://www.usenix.org/conference/osdi22/presentation/yu}.

\end{thebibliography}

\clearpage
\appendix
\section*{Supplementary Material}
\setcounter{table}{0}
\renewcommand{\thetable}{S\arabic{table}}
\renewcommand{\theHtable}{S\arabic{table}}

This supplement gives the configurations and checks supporting the main paper's precision, admission, replay, and cache comparisons. A cohort offers the 16 tasks together, with one agent session per task, at one setting, a choice of precision and admission limit $N$. The admission limit is the maximum number of agent sessions allowed to run concurrently. A paired repetition uses the same tasks, sampling seed, and admission order across the compared settings. BF16 uses 16-bit bfloat16 weights and activations, W8A8 uses 8-bit weights and activations, and W4A16 uses 4-bit weights with 16-bit activations. Energy per verified completion pools CPU-package plus GPU-board energy across a setting's cohorts, including failed attempts, and divides it by their independently verified task completions.

\section{Configuration and Workload}
\label{app:configuration}
We compare Devstral Small 1.1 on A100 PCIe deployments with 80 GB and 40 GB of memory, and Qwen3.5-9B on the 40 GB deployment. These configurations support the comparisons designed in Section 4 of the main paper. Table~\ref{tab:platform} gives Devstral's context limits and key--value (KV) cache capacities. Devstral runs vLLM 0.10.2~\cite{kwon2023pagedattention} with prefix caching enabled and 90\% of GPU memory reserved for serving.

\begin{table}[ht]
\centering\small
\caption{Devstral deployment settings and measured KV-cache capacity, in tokens. Both accelerators are A100 PCIe devices. Devstral BF16 fits only on the 80 GB deployment. Cache entries use BF16 throughout. The capped row sets both quantized precisions to BF16's capacity on 80 GB, and W4A16 to W8A8's capacity on 40 GB.}
\label{tab:platform}
\vskip 0.1in
\begin{tabularx}{\columnwidth}{lRR}
\toprule
\rowcolor{tableheadergray}
& \bfseries 80 GB & \bfseries 40 GB\\
\rowcolor{tableheadergray}
\bfseries Quantity & \bfseries deployment & \bfseries deployment\\
\midrule
Context limit (tokens) & 131,072 & 65,536\\
BF16 KV capacity & 176,368 & --\\
W8A8 KV capacity & 311,552 & 77,056\\
W4A16 KV capacity & 377,120 & 142,608\\
Capped KV capacity & 176,368 & 77,056\\
\bottomrule
\end{tabularx}
\end{table}

\paragraph{Single-agent comparisons.}
On 80 GB at $N=1$, every precision costs more per verified completion than in any of its tested shared settings. Over two cohorts per precision, energy per verified completion is 329.0 kJ for W4A16 and 400.7 kJ for BF16. Their ordering reverses at their lowest observed shared settings, where BF16 costs 98.1 kJ and W4A16 costs 114.3 kJ per verified completion.

\paragraph{Admission limits from the pilot runs.}
The admission grid in Table 1 of the main paper covers the pilot choices and the rules for low energy per output token and high token throughput. The pilot choices minimize energy per verified completion among settings meeting its completion and latency requirements, with the pilot cohorts flagged for tool timeouts either retained or excluded.

\paragraph{Devstral checkpoints and inference.}
We use the BF16 Devstral Small 1.1 checkpoint and the matched Red Hat W8A8 and W4A16 checkpoints~\cite{mistral2025devstral11,redhat2025devstralw4a16,redhat2025devstralw8a8}. W8A8 rescales weights and activations with SmoothQuant at strength 0.8, then quantizes weights with GPTQ~\cite{xiao2023smoothquant,frantar2023gptq}. It uses INT8 activations and the CUTLASS INT8 execution kernel. W4A16 uses asymmetric GPTQ with group size 128, BF16 activations, and the Marlin execution kernel. Sampling uses temperature 0.15 and top-$p=1$. Output length is bounded by the context limit and the session stopping rules.

\paragraph{Qwen3.5-9B.}
Qwen3.5-9B uses the publisher's BF16 checkpoint and matched Red Hat W8A8 and W4A16 checkpoints~\cite{qwen2026qwen35,redhat2026qwen35w4a16,redhat2026qwen35w8a8}. It runs vLLM 0.18.1 with 90\% of GPU memory reserved for serving. Both quantized checkpoints apply GPTQ and leave the linear-attention layers in 16-bit precision. Sampling follows the model card's non-thinking settings (temperature 0.7, top-$p=0.8$, top-$k=20$, min-$p=0$, presence penalty 1.5, and repetition penalty 1.0), with thinking disabled in the chat template. The measured KV-cache capacities are 149,424, 189,552, and 209,088 tokens for BF16, W8A8, and W4A16, respectively. W8A8 uses the CUTLASS INT8 execution kernel, and W4A16 uses Marlin.

\begin{table*}[t]
\centering\small
\caption{Task catalogue, difficulty, and measured turns and output tokens. The upper block lists data tasks and input sizes, and the lower block lists repository issues and SWE-bench Verified's human estimates of an engineer's time~\cite{jimenez2024swebench,openai2024swebenchverified}. One turn is one model response. Turns and output tokens are medians over six repetitions of BF16 with four admitted agents on 80 GB. Passes count verified completions out of six attempts per task.}
\label{tab:tasks}
\vskip 0.1in
\begin{tabularx}{\textwidth}{llRRR}
\toprule
\rowcolor{tableheadergray}
& \bfseries Input size or & & \bfseries Output & \\
\rowcolor{tableheadergray}
\bfseries Task & \bfseries estimated repair time & \bfseries Turns & \bfseries tokens & \bfseries Passes\\
\midrule
SQL analytics, segments & 8,192 rows & 5 & 552 & 6/6\\
SQL analytics, declines & 1,000,000 rows & 4.5 & 723.5 & 6/6\\
Grouped regression & 8,192 rows & 3 & 630.5 & 6/6\\
Grouped regression & 1,000,000 rows & 3 & 783 & 6/6\\
ETL repair & 8,192 rows & 4 & 384.5 & 4/6\\
ETL repair & 1,000,000 rows & 4 & 378 & 4/6\\
Python, deduplication & 8,192 rows & 4 & 215 & 6/6\\
Python, histogram & 100,000 rows & 4 & 266.5 & 6/6\\
\midrule
Django 16801 & $<$15 min & 18 & 1,518 & 5/6\\
Django 16899 & $<$15 min & 170 & 22,068 & 0/6\\
Sphinx 9258 & $<$15 min & 73 & 11,410.5 & 0/6\\
Matplotlib 21568 & 15 min--1 h & 119.5 & 21,044 & 0/6\\
Sphinx 8120 & 15 min--1 h & 35 & 12,798.5 & 1/6\\
SymPy 19783 & 15 min--1 h & 24.5 & 4,251 & 0/6\\
scikit-learn 25102 & 1--4 h & 9.5 & 10,816.5 & 0/6\\
SymPy 13852 & 1--4 h & 41.5 & 6,253 & 0/6\\
\bottomrule
\end{tabularx}
\end{table*}

\paragraph{Qwen3.5-9B validation.}
Before measurement, we checked prompt rendering, valid and invalid tool calls, and exact replay. A BF16 screen on eight tasks outside the catalogue required at least three verified completions, including one repository issue, and passed with six. All 33 four-agent cohorts and eight 16-agent cohorts pass the inclusion checks in Section 4 of the main paper. The eight 16-agent cohorts end before the watchdog, which stops a session after 1,800 seconds. The 11 four-agent repetitions use fresh task orders and seeds and are pooled under a rule fixed before outcomes were examined. Repetitions 0 and 3 include duplicate replays of each recording on the same precision. Their energies differ by at most 0.6\%, and we average them. Four paired repetitions at $N=16$ use fresh task orders and seeds. Their design and analysis were fixed in advance to test the prediction that W4A16 would remain costlier per verified completion. The resulting ordering is unresolved (Section 5.3 of the main paper).

\paragraph{Verification and measurement boundary.}
Independent verification runs outside the energy interval and has a separate 900-second timeout. Repository containers receive four CPUs and 16 GB of memory. Cohort energy includes failed attempts and excludes model loading and cache reset.

\paragraph{Task catalogue.}
Table~\ref{tab:tasks} gives each task's scale, difficulty, and measured work. Across BF16's 48 repository attempts with four admitted agents on 80 GB, the median is 31.5 turns, and 10 attempts reach the 250-turn budget. BF16 verifies 6 of these 48 repository attempts. All precision comparisons use the same minimal tool-calling loop. Devstral's published 53.6\% completion rate on SWE-bench Verified uses the OpenHands scaffold~\cite{mistral2025devstral11}, while our catalogue includes eight selected issues, two rated 1--4 h.

\paragraph{Measurement and uncertainty.}
Cohort energy sums the two CPU-package counters and the serving GPU's board-energy counter from the NVIDIA Management Library (NVML). Core subdomains are already included in the package counters. We sample energy at a nominal 10 Hz and telemetry at 1 Hz, retain each reading's actual timestamp, and correct counter wraps (overflows back to zero). The CPU packages and GPU board account for part of the machine's wall-plug energy. Monitoring runs in every cohort, so its cost is included in every total.

We obtain intervals by resampling whole paired repetitions 10,000 times. Each draw uses the same repetitions for the compared settings and recomputes each setting's summed energy over summed successes. Devstral's primary comparisons use paired differences in energy per verified completion. The Qwen3.5-9B comparison at $N=16$ uses a prespecified ratio. Ratio intervals throughout the main paper use the same paired resampling. Every draw in the reported Devstral comparisons retains a success in each setting, so each ratio is defined.

Exact paired sign-flip tests use per-repetition cost differences, whereas the bootstrap intervals recompute pooled costs. The two procedures can give different decisions at a 5\% threshold. For Devstral's W4A16 versus W8A8, the two-sided tests give $p=0.063$ at $N=4$ on 80 GB (six pairs) and $p=0.031$ at both tested limits on 40 GB (seven pairs). For Qwen3.5-9B, they give $p=0.002$ at $N=4$ (11 pairs) and $p=0.75$ at $N=16$ (four pairs). The exploratory 40 GB cache cap at $N=16$ gives $p=0.25$ over four pairs, with three showing an increase in cost.

\section{Validity and Sensitivity}
\label{app:validity}
The checks below support the live comparisons, replay of recorded calls, and cache controls in Section 5 of the main paper. In the main 80 GB experiment, the 40 GB Devstral cohorts with and without the cache cap, and the Qwen3.5-9B cohorts, every verified completion we examined finishes within 649 seconds of service. The 1,800-second watchdog allows substantially more time than these observed successes. Of the stopped sessions, 86\% on 80 GB (98 of 114) and 70\% on 40 GB (124 of 178) spend their last 300 seconds repeating earlier calls, sending malformed calls, or generating repetitive text. The 40 GB stops also include 19 W8A8 sessions with 16 admitted agents still waiting for a first token under cache pressure.
\paragraph{Retained and incomplete measurements.}
All 54 main 80 GB cohorts account for all 16 tasks and have valid measured energy. In each cohort, all pending server requests finish (server drain), and we retain all 54 cohorts. Four cohorts flagged for tool errors contain only watchdog messages, which count as task outcomes under the classification fixed before the main experiment.

The 40 GB primary comparison uses the seven complete paired repetitions, 0--2 and 4--7. An infrastructure failure left repetition 3 incomplete while saving the W4A16 agent's code changes at $N=4$ as a patch containing non-UTF-8 bytes. From repetition 4 onward, the saving procedure preserves binary patches for grading. The primary comparison retains complete pairs from before and after that revision.

W4A16 remains cheaper in the later 40 GB repetitions, supporting the primary precision ordering. An operating-system library update occurred during the early runs, and the primary comparison retains the affected repetition's complete pair. Repetitions 4--11, after the patch-saving revision, give W4A16/W8A8 energy-per-verified-completion ratios of 0.571 at $N=4$ (95\% interval $[0.407,0.771]$) and 0.602 at $N=16$ ($[0.407,0.810]$). These eight pairs include four primary repetitions and the normal-capacity cohorts of the four cache-cap repetitions.

\paragraph{Checks on exported measurements.}
Recounting the verifier records reproduces the success counts of all 54 main, six single-agent, and 16 equal-cache cohorts on 80 GB. Exports from the 80 GB main and equal-cache comparisons and the 40 GB Devstral runs match their recorded file checksums. Integrating the raw counter traces reproduces every cohort's energy summary within the prespecified 0.5\% tolerance.

\paragraph{Fidelity and energy of crossed replay.}
Replay executes recorded model calls across precisions while holding their prompt and output token IDs fixed. All nine Devstral replays complete every recorded session, reproduce every output-token sequence, and drain the server, so Table~\ref{tab:replay} compares the energy of executing the same recorded work. For Qwen3.5-9B, the replay crosses W8A8 and W4A16. All 52 Qwen3.5-9B replays reproduce every recorded turn exactly.

\begin{table}[ht]
\centering\small
\caption{W4A16 executes every recorded workload with the least energy, and the workload recorded under W4A16 uses the most on every executing precision. Crossed-replay energy (kJ) on 80 GB, one replay per value. Rows identify the precision that recorded the $N=4$ workload, and columns identify the executing precision. The last column gives the within-row W4A16/W8A8 ratio using unrounded energies.}
\label{tab:replay}
\vskip 0.1in
\begin{tabularx}{\columnwidth}{lRRRR}
\toprule
\rowcolor{tableheadergray}
\bfseries Recorded & \multicolumn{3}{c}{\bfseries Executing precision} & \bfseries W4A16/\\
\rowcolor{tableheadergray}
\bfseries workload & \bfseries BF16 & \bfseries W8A8 & \bfseries W4A16 & \bfseries W8A8\\
\midrule
BF16 & 903 & 643 & 485 & 0.754\\
W8A8 & 609 & 433 & 326 & 0.754\\
W4A16 & 1,222 & 901 & 695 & 0.771\\
\bottomrule
\end{tabularx}
\end{table}

\paragraph{Equal-cache pressure.}
The 80 GB equal-cache comparison leaves the precision ordering unresolved over four paired repetitions. Both quantized precisions receive BF16's capacity of 176,368 tokens. The W4A16-minus-W8A8 difference is $-46.6$ kJ per verified completion at $N=4$ (95\% interval $[-126.5,13.6]$) and $11.3$ kJ per verified completion at $N=16$ ($[-35.5,45.7]$). Both intervals include zero. Repetition 0 is the original pass, and the three added repetitions reuse the task orders and sampling seeds of main repetitions 1--3. Their number and analysis were fixed before the main 80 GB outcomes were inspected. These added repetitions ran later than the normal-capacity cohorts, so a comparison with those cohorts can also reflect changes over time. Table~\ref{tab:pressure} shows how waiting and preemptions vary across the capped cohorts. A preemption interrupts a request under cache pressure so that its state can be recovered later.

\begin{table}[ht]
\centering\small
\caption{Cache pressure under the cap varies across repetitions. Waiting share (percent)/preemptions in each 80 GB cohort capped at BF16's cache capacity. Waiting share is the fraction of active telemetry samples with at least one request waiting. Column 0 is the original pass, and columns 1--3 are the three added repetitions.}
\label{tab:pressure}
\vskip 0.1in
\setlength{\tabcolsep}{3pt}
\begin{tabularx}{\columnwidth}{lRRRR}
\toprule
\rowcolor{tableheadergray}
& \multicolumn{4}{c}{\bfseries Repetition}\\
\rowcolor{tableheadergray}
\bfseries Setting & \bfseries 0 & \bfseries 1 & \bfseries 2 & \bfseries 3\\
\midrule
W4A16, $N=4$ & 29.34/1 & 0.06/0 & 0.65/0 & 0.04/0\\
W8A8, $N=4$ & 28.26/16 & 0/0 & 0/0 & 42.43/5\\
W4A16, $N=16$ & 76.94/36 & 69.63/7 & 0.33/0 & 0.11/0\\
W8A8, $N=16$ & 0.06/0 & 0.06/0 & 4.31/0 & 22.11/1\\
\bottomrule
\end{tabularx}
\end{table}

In the original pass, capping the cache increases both waiting and preemptions relative to the corresponding normal-capacity cohorts in three of four settings, while W8A8 at $N=16$ stays close to its normal-capacity level. Using only the three added repetitions, the difference (W4A16 minus W8A8) is $-18.9$ kJ per verified completion at $N=4$ (95\% interval $[-93.0,31.0]$) and is $4.4$ kJ per verified completion at $N=16$ ($[-60.3,47.8]$). Both intervals include zero, so the added repetitions alone leave the order of the precisions open.

\section{Latency and the Final Session}
\label{app:support}

The latency comparison in Section 5.4 of the main paper uses service time from admission to session end. Table~\ref{tab:latency} also reports queue-inclusive time, which adds waiting for admission. At $N=4$, this waiting lengthens the time a task spends in the service. At $N=16$, all tasks are admitted together, so the two times nearly coincide. Session end includes watchdog cancellation and cleanup, which are included in the 1,802-second p90 values. The cohort drain that follows session end remains part of the energy interval.

\begin{table}[ht]
\centering\small
\caption{80 GB latency in seconds over 96 offered task attempts for each of the six settings in Table 2 of the main paper. The p50 and p90 columns give the median and 90th percentile, using nearest-rank quantiles. Service time starts at admission, and queue-inclusive time starts when all tasks are offered.}
\label{tab:latency}
\vskip 0.1in
\begin{tabularx}{\columnwidth}{lRRRR}
\toprule
\rowcolor{tableheadergray}
& \multicolumn{2}{c}{\bfseries Service} & \multicolumn{2}{c}{\bfseries Queue-inclusive}\\
\rowcolor{tableheadergray}
\bfseries Setting & \bfseries p50 & \bfseries p90 & \bfseries p50 & \bfseries p90\\
\midrule
BF16, $N=4$ & 43 & 1,594 & 270 & 1,834\\
W8A8, $N=4$ & 35 & 858 & 262 & 1,380\\
W4A16, $N=4$ & 43 & 1,802 & 149 & 1,830\\
BF16, $N=16$ & 73 & 1,802 & 73 & 1,802\\
W8A8, $N=16$ & 99 & 1,802 & 99 & 1,802\\
W4A16, $N=16$ & 91 & 1,802 & 92 & 1,802\\
\bottomrule
\end{tabularx}
\end{table}

\paragraph{Energy after other sessions have finished.}
A final session can account for a large share of a cohort's energy, so the measurement includes the period after all other sessions have finished (Section 3 of the main paper). This period begins when one session runs alone with an empty admission queue. Its median shares of main-cohort energy are 0.79\% for BF16, 4.55\% for W8A8, and 3.67\% for W4A16. Maximum shares are 41.4\%, 68.4\%, and 44.4\%, respectively.

\end{document}